\documentstyle[aps,epsf,epsfig]{revtex}

\newcommand{\bq}{\begin{equation}}
\newcommand{\eq}{\end{equation}}
\newcommand{\bqn}{\begin{eqnarray}}
\newcommand{\eqn}{\end{eqnarray}}
\newcommand{\nb}{\nonumber}
\newcommand{\lb}{\label}

\begin{document}
\title{Collapsing Scalar Field with Kinematic Self-Similarity of the Second
Kind in $2+1$ Gravity}
\author{R. Chan\thanks{E-mail: chan@on.br} $^1$,
M. F. A. da Silva\thanks{E-mail: mfas@dtf.if.uerj.br} $^2$ , 
J. F. Villas da Rocha\thanks{E-mail: roch@dft.if.uerj.br} $^2$ ,
  and Anzhong Wang\thanks{E-mail: Anzhong$\_$Wang@baylor.edu}
  $^{2,3}$}
\address{
$^1$ Coordenadoria de Astronomia e Astrof\'{\i}sica,
 Observat\'orio Nacional, Rua General Jos\'e Cristino 77,
S\~ao Crist\'ov\~ao,
CEP. $20921-400$, Rio de Janeiro, RJ, Brazil \\
$^2$ Departamento de F\'{\i}sica Te\'orica, Universidade
do Estado do Rio de Janeiro, 
Rua S\~ao Francisco Xavier $524$, Maracan\~ a,
CEP. $20550-013$, Rio de Janeiro, RJ, Brazil \\
$^3$ CASPER, Department of Physics, P.O. Box 97316, 
Baylor University, Waco, TX76798-7316 - USA}

\date{\today}

\maketitle

\begin{abstract}

All the $2+1$-dimensional circularly symmetric solutions with
kinematic self-similarity of the second kind to the
Einstein-massless-scalar field equations are found and their local
and global properties are studied. It is found that some of them
represent gravitational collapse of a massless scalar field, in
which  black holes are always formed.

\end{abstract}

\vspace{.6cm}


\section{Introduction}

The studies of non-linearity of the Einstein field equations near
the threshold
  of
black hole formation reveal very rich phenomena \cite{Chop93},
which are quite similar to critical phenomena in Statistical
Mechanics and Quantum Field Theory \cite{Golden}. In particular,
by numerically studying the gravitational collapse of a massless
scalar field in $3+1$-dimensional spherically symmetric
spacetimes, Choptuik found that the mass of such formed black
holes takes a scaling form,
  \bq
  \lb{1.1}
  M_{BH} = C(p)\left(p -p^{*}\right)^{\gamma},
  \eq
where $C(p)$ is a constant and depends on the initial data, and
$p$ parameterizes a family of initial data in such a way that when
$p > p^{*}$  black holes are formed, and when $p < p^{*}$ no black
holes are formed. It was shown that, in contrast to $C(p)$, the
exponent $\gamma$   is universal to all the families of initial
data studied. Numerically it was determined as $\gamma \sim 0.37$.
The solution with $p = p^{*}$, usually called the critical
solution, is found also universal. Moreover, for the massless
scalar field it is periodic, too. Universality  of the critical
solution and exponent, as well as the power-law scaling of the
black hole mass all have given rise to the name {\em Critical
Phenomena in Gravitational Collapse}. Choptuik's studies were soon
generalized to other matter fields \cite{Gun00,Wang01}, and now
the following seems clear: (a) There are two types of critical
collapse, depending on whether the black hole mass takes the
scaling form (\ref{1.1}) or not. When it takes the scaling form,
the corresponding collapse is called Type $II$ collapse, and when
it does not it is called Type $I$ collapse. In the type $II$
collapse, all the critical solutions found so far have either
discrete self-similarity (DSS) or homothetic self-similarity
(HSS), depending on the matter fields. In the type $I$ collapse,
the critical solutions have neither DSS nor HSS. For certain
matter fields, these two types of collapse can co-exist. (b) For
Type $II$ collapse, the corresponding exponent is universal only
with respect to certain matter fields. Usually, different matter
fields have different critical solutions and, in the sequel,
different exponents. But for a given matter field the critical
solution and the exponent are universal \footnote{So far, the
studies have been mainly restricted to spherically symmetric case
and their non-spherical linear perturbations. Therefore, it is not
really clear whether or not the critical solution and exponent are
universal with respect to different symmetries of the
spacetimes \cite{Cho03,Wang03}.}. (c)  A critical solution for both of the two
types has one and only one unstable mode. This now is considered as one
of the main criteria for a solution to be critical. (d) The
universality of the exponent is closely  related to the last
property. In fact, using dimensional analysis \cite{Even} one can
show that
  \bq
  \lb{1.2}
  \gamma = \frac{1}{\left|k_{1}\right|},
  \eq
where $k_{1}$ denotes the unstable mode.

From the above, one can see that to study (Type $II$)
critical collapse, one
may first find some particular solutions by imposing certain
symmetries, such as, DSS or HSS. Usually this considerably
simplifies the problem. For example, in the spherically symmetric
case, by imposing HSS symmetry the Einstein field equations can be
reduced from PDE's to ODE's.    Once the particular solutions are
known, one can study their linear perturbations and find out the
spectrum of the corresponding eigen-modes. If a solution has one
and only one unstable mode, by definition we may consider it as a
critical solution (See also the discussions given in
\cite{Brady02}). The studies of critical collapse  have been
mainly numerical so far, and analytical ones are still highly
hindered by the complexity of the problem, even after imposing
some symmetries.

Lately, Pretorius and Choptuik (PC) \cite{PC00} studied
gravitational collapse of a massless scalar field in an anti-de
Sitter background in $2+1$-dimensional spacetimes with circular
symmetry, and found that the  collapse exhibits  critical
phenomena and the mass of such formed black holes takes the
scaling form of Eq.(\ref{2.1}) with $\gamma = 1.2 \pm 0.02$, which
is different from that of the corresponding $3+1$-dimensional
case.  In addition, the critical solution is also different, and,
instead of having DSS, now has HSS. The above results were
confirmed by independent numerical studies \cite{HO01}. However,
the exponent obtained by Husain and Olivier (HO),  $\gamma \sim
0.81$,  is quite different from the one obtained by  PC. It is not
clear whether the difference is due to numerical errors or to some
unknown physics.

After the above numerical work, analytical studies of the same
problem soon followed up \cite{Gar01,CF01,GG02,HW02}. In
particular, Garfinkle found a class, say, $S[n]$, of exact
solutions to the Einstein-massless-scalar field equations and
showed that in the strong field regime the $n = 4$ solution fits
very well with the numerical critical solution found by PC.
Lately, Garfinkle and Gundlach (GG) studied their linear
perturbations and found that only the solution with $n = 2$ has
one unstable mode, while the one with $n = 4$ has three
\cite{GG02}. According to  Eq.(\ref{2.2}), the corresponding
exponent is given by $\gamma = 1/|k_{1}| = 4/3$. Independently,
Hirschmann, Wu and one of the present author (HWW) systematically
studied the problem, and  found that the $n = 4$ solution indeed
has only one unstable mode \cite{HW02}. This difference actually
comes from the use of different boundary conditions. As a matter
of fact, in addition to the ones imposed by GG \cite{GG02}, HWW
further required that no matter field should come out of the
already formed black holes. This additional condition seems
physically quite reasonable and has been widely used in the
studies of black hole perturbations \cite{Chandra83}. However, now
the corresponding exponent is given by $\gamma = 1/|k_{1}| = 4$,
which is significantly different from the numerical ones. So far,
no  explanations about these differences have been worked out,
yet.

In this paper we do not intend to solve the above problems, but
study another class of exact solutions with kinematic
self-similarity of the second kind. Self-similarity is usually
divided into two classes, one is the discrete self-similarity
mentioned above, and the other is the so-called kinematic
self-similarity (KSS) \cite{CH89}, and sometimes it is also called
continuous self-similarity (CSS). KSS or CSS is further classified
into three different kinds, the zeroth, first and second. The
kinematic self-similarity of the first kind is also called
homothetic self-similarity, first introduced to General Relativity
by Cahill and Taub in 1971 \cite{CT71}. In Statistical Mechanics,
critical solutions with KSS of the second kind seem more generic
than those of the first kind \cite{Golden}. However,  critical
solutions with KSS of the second kind have not been found so far
in gravitational collapse, and it would be very interesting to
look for such solutions.  In this paper we shall present all the
solutions of the Einstein-massless scalar field equations in $2+1$
dimensional circularly symmetric spacetimes with KSS of the second
kind, and then study their local and global properties. The study
of their linear perturbations will be considered somewhere else.

\section{Massless Scalar Field with Kinematic Self-similarity of the
          Second Kind}

The $2+1$-dimensional spacetimes with circular symmetry are
described by the metric
  \bq
  \lb{2.1}
  ds^{2} = \gamma_{ab}\left(t, r\right) dx^{a}dx^{b}
  + g_{\theta\theta}\left(t, r\right) d\theta^{2},
  \eq
where $\left\{x^{a}\right\} = \left\{t, r\right\}, \; (a, b = 0,
1)$, and $\theta$ denotes the angular coordinate, with the
hypersurfaces $\theta = 0, \; 2\pi$ being identified. Clearly, the
metric is invariant under the following coordinate
transformations,
  \bq
  \lb{2.2}
   t = t\left(t', r'\right),\;\;\;
   r = r\left(t', r'\right).
  \eq
On the other hand, for a massless scalar field $\phi$, using the
Einstein field equations
  \bq
  \lb{2.3}
  R_{\mu\nu} = \kappa\phi_{,\mu}\phi_{,\nu},
  \eq
  it can be shown that   the scalar field $\phi$ in general is
function of $t$ and $r$, where $(\;)_{,\mu} \equiv
\partial(\;)/\partial x^{\mu}$ and $\kappa $ denotes the Einstein
coupling constant. The Greek letters run from $ 0$ to $2$. In this
paper we shall choose units such that $\kappa = 1$. Considering
the fact that a collapsing scalar field must be timelike,  using
the gauge freedom (\ref{2.2}), we shall choose, without loss of
any generality, the coordinates such that
  \bq
   \lb{gauge}
   g_{01} \left(t, r\right) = 0, \;\;\;\;
   \phi\left(t, r\right) = 2q\ln(-t),
   \eq
where $q$ is a constant. When $q = 0$,   the spacetime is vacuum
and then must be flat \cite{Carlip}. Therefore, in the following
we shall assume that $q \not= 0$. It is interesting to note that
the last expression in Eq.(\ref{gauge})  is physically equivalent
to choose the coordinates such that they are comoving with the
scalar field. Since  in the present case $\phi_{,\mu}$ is
timelike, clearly   this is always possible. The gauge given by
Eq.(\ref{gauge}) will be called {\em comoving gauge}, for which
the metric (\ref{2.1}) can be cast in the form
  \bq
  \lb{metric}
    ds^{2} = l^{2}\left\{e^{2\Phi(t, r)}dt^{2} - e^{2\Psi(t, r)}dr^{2}
             - r^{2}S^{2}(t, r)d\theta^{2}\right\},
  \eq
where  $l$ is an unit constant with the dimension of length, so
that all the coordinates $\left\{x^{\mu}\right\} = \left\{t, r,
\theta\right\}$ are dimensionless. From $\phi$ we can construct a
timelike unit vector $u_{\mu} $  and a projector operator
$h_{\mu\nu}$ by
  \bqn
   \lb{2.4}
   u_{\mu} &\equiv&
\frac{\phi_{,\mu}}{\left(\phi_{,\alpha}\phi^{,\alpha}\right)^{1/2}}
  = le^{\Phi}\delta^{0}_{\mu},\nb\\
h_{\mu\nu} &\equiv& g_{\mu\nu} - u_{\mu}u_{\nu},
  \eqn
from which we find $ h^{\alpha\beta} u_{\alpha}u_{\beta} = 0$.
Once the project operator $h_{\mu\nu}$ is defined, Following
Carter and Henriksen \cite{CH89}, we define kinematic
self-similarity by
  \bq
  \lb{2.5}
  {\cal{L}}_{\xi}h_{\mu\nu} = 2 h_{\mu\nu},\;\;\;\;
  {\cal{L}}_{\xi}u^{\mu} = -\alpha u^{\mu},
  \eq
where ${\cal{L}}_{\xi}$ denotes the Lie differentiation along the
vector field $\xi^{\mu}$, $\alpha$ is a {\em dimensionless}
constant. When $\alpha = 0$, the corresponding solutions are said
to have self-similarity of {\em the zeroth kind}, when $\alpha =
1$, they are said to have  self-similarity of {\em the first kind}
(or {\em homothetic self-similarity}), and when $\alpha \not= 0,
1$, they are said to have self-similarity of {\em the second
kind}.

All the solutions with KSS of the first kind were given in
\cite{HW02}. Also, it can be shown that {\em no solutions with KSS
of the zeroth kind to the Einstein-massless scalar field equations
with circular symmetry exist}. Thus, in the following we shall
consider only solutions with KSS of the second kind ($\alpha
\not=0, 1$). Applying the above definition to metric
(\ref{metric}), we find that
  \bq
  \lb{SS}
  \Phi(t,r) = \Phi(x),\;\;\;
  \Psi(t,r) = \Psi(x),\;\;\;
  S(t,r) = S(x),
  \eq
where the self-similar variable $x$ and the vector field
$\xi^{\mu}$ are given by
  \bqn
  \lb{2.7}
  & & \xi^{\mu}\frac{\partial}{\partial x^{\mu}}
    = \alpha t \frac{\partial}{\partial t} + r\frac{\partial}
    {\partial r},\nb\\
  & & x = \ln (r) - \frac{1}{\alpha}\ln(-t),\;\; (\alpha \not= 0, 1).
  \eqn

Before looking for the solutions of the Einstein field equations,
we would like to note that for the metric (\ref{metric})  to
represent circular symmetry, some physical and geometrical
conditions needed to be imposed \cite{Fatima}. For gravitational
collapse, we impose the following conditions:

(i) There must exist a symmetry axis, which can be expressed as
   \bq
   \lb{cd1}
   {\cal{R}} \equiv \left|\xi^{\mu}_{(\theta)}\xi^{\nu}_{(\theta)}
      g_{\mu\nu} \right|^{1/2} \rightarrow 0,
   \eq
as $r \rightarrow 0$, where we have chosen the radial coordinate
such that the axis is located at $r = 0$, and
$\xi^{\mu}_{(\theta)}$ is the Killing vector with a close orbit,
and given by $\xi^{\alpha}_{(\theta)}\partial_{\alpha} =
\partial_{\theta}$.

(ii) The spacetime near the symmetry axis is locally flat, which
can be written as \cite{Kramer80}
  \bq
  \lb{cd2}
  {\cal{R}}_{,\alpha}{\cal{R}}_{,\beta} g^{\alpha\beta}
  \rightarrow - 1,
  \eq
as  $r \rightarrow 0$. Note that solutions failing to satisfy this
condition   sometimes are also acceptable. For example, when the
right-hand side of the above equation approaches a finite constant,
the singularity at $r = 0$ may be related to a point-like particle
\cite{VS}. However, since here we are mainly interested in
gravitational collapse, in this paper we shall assume that this
condition holds strictly at the beginning of the collapse, so that
we can be sure that the singularity to be formed later on the axis
is due to the collapse.

(iii) No closed timelike curves (CTC's). In spacetimes with
circular symmetry, CTC's can be easily introduced. To ensure
  their  absence, we assume that the condition
  \bq
  \lb{cd3}
\xi^{\mu}_{(\theta)}\xi^{\nu}_{(\theta)}g_{\mu\nu} < 0,
  \eq
holds in the whole spacetime.

In addition to these conditions, it is usually also required that
the spacetime be asymptotically flat in the radial direction.
However, since we consider solutions with self-similarity, this
condition cannot be satisfied, unless we restrict the validity of
them only up to a maximal radius, say, $r = r_{0}(t)$, and then
join them with others in the region $r > r_{0}(t)$, which are
asymptotically flat as $r \rightarrow \infty$. In this paper, we
shall not consider such a possibility, and simply assume that the
self-similar solutions are valid in the whole spacetime.

It should be noted that the gauge conditions (\ref{gauge}), the
self-similarity conditions (\ref{SS}) and the regularity
conditions (\ref{cd1})-(\ref{cd3}) do not completely fix the gauge
freedom. As a matter of fact, it can be shown that the metric and
all these conditions are invariant under the coordinate
transformations
  \bq
  \lb{rescaling}
  t = A \bar{t},\;\;\; r = B \bar{r},
  \eq
where $A$ and $B$  are arbitrary constants. Using this remaining
gauge freedom, we shall further assume that
  \bq
  \lb{cd4}
  \Phi(t, 0) = 0,
  \eq
that is, the timelike coordinate $t$ measures the proper time on
the axis.

Substituting Eqs.(\ref{gauge}) and (\ref{SS}) into the Einstein
field equations (\ref{2.3}), we find the following equations
  \bqn
  \lb{eq1}
  y_{,x} - (1+y)\left(\Psi_{,x} - y\right) - y\Phi_{,x} &=& 0,\\
  \lb{eq2}
  \Phi _{,xx}+\Phi _{,x}\left( \Phi _{,x}-\Psi _{,x}-y-2\right) &=&0,\\
  \lb{eq3}
  \Psi _{,xx} - \Psi _{,x}\left(\Phi _{,x}-\Psi _{,x} + y + 1-\alpha \right)
  + \left(1 - \alpha\right)y  &=&0,\\
  \lb{eq4}
   \left(1 + 2y\right)\Phi_{,x} &=& 0,\\
  \lb{eq5}
   \left(1 + 2y\right)\Psi_{,x} - (1-\alpha)y &=& 0,\\
  \lb{eq6}
   y\Phi_{,x} &=& 0,\\
  \lb{eq7}
   y\Psi_{,x} - 2\alpha^{2}q^{2} &=&0,
  \eqn
where $y \equiv S_{,x}/{S}$. Since $\alpha \not=0$ and $q \not=0$,
from Eqs.(\ref{eq4})-(\ref{eq7}) we find that
  \bq
  \lb{2.8a}
  \Phi_{,x} = 0, \;\;\;\;  \Psi_{,x} \not= 0, \;\;\;
  y \not= 0,\; - \frac{1}{2}.
  \eq
To study the above equations further, it is found convenient to
consider the two cases $y \not= -1$ and $y = -1$ separately.

When $y \not= -1$, differentiating Eq.(\ref{eq5}) we find that
  \bq
\lb{2.9a}
  \Psi_{,xx} = \frac{y_{,x}}{1 + 2y}\left[(1- \alpha) -
2\Psi_{,x}\right],
  \eq
while from Eq.(\ref{eq1}) we obtain
  \bq
  \lb{2.9b}
  \Psi_{,x} = \frac{y_{,x}}{1+y} + y.
  \eq
Inserting Eqs.(\ref{2.9a}) and (\ref{2.9b}) into Eq.(\ref{eq3})
and considering Eq.(\ref{2.8a}), we obtain
  \bq
  \lb{2.10}
  y_{,x}\left[y_{,x} +
  \left(2-\alpha\right)y\left(1+y\right)\right] = 0.
  \eq
Thus, there are two possibilities,
  \bqn
  \lb{2.11a}
  & & (i) \;\; y =  y_{0},\\
\lb{2.11b} & & (ii) \;\; y_{,x} +
\left(2-\alpha\right)y\left(1+y\right) = 0,
  \eqn
where $y_{0}$ is a constant. When $y = y_{0}$, the integration of
Eq.(\ref{2.9b}) yields
  \bq
\lb{2.12}
  \Psi(x) = y_{0} x + \Psi_{0},
  \eq
where $\Psi_{0}$ is an integration constant. On the other hand,
from Eq.(\ref{eq5}) we find that $y_{0} = - \alpha/2$, while
Eq.(\ref{eq7}) gives $q^{2} = 1/8$. It can be shown that all the
other equations are satisfied identically. Thus, in this case we
have the following solutions
  \bqn
  \lb{class1}
  \Phi(x) &=& \Phi_{0},\;\;\;
  S(x) = S_{0}e^{-\alpha x/2},\nb\\
  \Psi(x) &=& -\frac{1}{2}\alpha x + \Psi_{0}, \;\;\;
  q =   \pm \frac{1}{\sqrt{8}},
  \eqn
where $\Phi_{0}$ and $S_{0}$ are other integration constants.

When Eq.(\ref{2.11b}) holds, combining it with Eq.(\ref{eq1}) we
find that $\Psi_{,x} = - (1-\alpha)y$. Then, substituting it into
Eq.(\ref{eq5}) we obtain $y + 1 = 0$. Since in the present case we
assume that $y \not= -1$, we can see that in this case there is no
solution.

When $y = -1$, Eq.(\ref{eq5}) yields
  \bq
   \lb{2.13}
   \Psi_{,x} = -(1-\alpha)y.
   \eq
Inserting it into Eq.(\ref{eq7}) we obtain $q^{2} = (\alpha -
1)/(2\alpha^{2})$, while all the other Einstein field equations
are satisfied identically. Thus, now the general solutions are
given by
  \bqn
  \lb{class2}
\Phi(x) &=& \Phi_{0},\;\;\; S(x) = S_{0}e^{-x},\nb\\
\Psi(x) &=&  (1-\alpha) x + \Psi_{0}, \;\;\;
  q = \pm \left(\frac{\alpha - 1}{2\alpha^{2}}\right)^{1/2}.
  \eqn
Clearly, to have the scalar field be real, now we must assume
$\alpha > 1 $.

\section{Local and Global Properties of the Self-Similar solutions}

To study the solutions found in the last section, following
\cite{HW02} let us first calculate the expansions of radially out-
and in-going null geodesics. To this end, we first introduce two
null coordinates $u$ and $v$ via the relations \cite{Yasuda}
   \bq
   \lb{3.1}
   du = f\left(e^{\Phi} dt - e^{\Psi}dr\right),\;\;\;
   dv = g\left(e^{\Phi} dt + e^{\Psi}dr\right),
  \eq
where $f$ and $g$ satisfy the integrability conditions for $u$ and
$v$, and, without loss of generality, we shall assume that
  $f > 0$ and $ g > 0$. Then, we can see that  the rays moving along the
  hypersurfaces $u = constant$ are outgoing, while the ones moving along the
hypersurfaces $v = constant$ are ingoing. The expansions of these
null geodesics are defined as \cite{HW02},
  \bqn
   \lb{3.2}
\theta_{l} &\equiv& \nabla_{\lambda}{l^{\lambda}}
   = e^{-2\sigma}\frac{{\cal{R}}_{,v}}{l^{2}{\cal{R}}}
   = \frac{f}{l^{2}{\cal{R}}}\left(e^{-\Phi}{\cal{R}}_{,t}
   + e^{-\Psi}{\cal{R}}_{,r}\right),\nb\\
   \theta_{n} &\equiv& \nabla_{\lambda}{n^{\lambda}}
   = e^{-2\sigma}\frac{{\cal{R}}_{,u}}{l^{2}{\cal{R}}}
   = \frac{g}{l^{2}{\cal{R}}}\left(e^{-\Phi}{\cal{R}}_{,t}
   - e^{-\Psi}{\cal{R}}_{,r}\right),
   \eqn
where $\nabla_{\lambda}$ denotes the covariant derivative,
${\cal{R}} = rS(x)$ and $\sigma \equiv -(1/2)\ln(fg)$. The null
vector $l^{\lambda}$ ($n^{\lambda}$) defines the outgoing
(ingoing) null geodesics, given by $ l_{\lambda}   =
  \delta^{u}_{\lambda}$ ($ n_{\lambda}   =\delta^{v}_{\lambda}$).
In terms of the self-similar variable $x$, Eq.(\ref{3.2}) becomes
  \bqn
  \lb{3.3}
  \theta_{l} &=& \frac{f}{\alpha l^{2}r}\left\{\alpha\left(1+y\right)
               e^{-\Psi}
               + y e^{x + (\alpha -1)\tau/\alpha - \Phi}\right\},\nb\\
   \theta_{n} &=& - \frac{g}{\alpha l^{2}r}\left\{\alpha\left(1+y\right)
              e^{-\Psi}
               - y e^{x + (\alpha -1)\tau/\alpha - \Phi}\right\},
               \;\;(\alpha \not= 0).
  \eqn
The apparent horizon is defined as the outmost surface of
$\theta_{l} = 0$ \cite{Hay94}. In the
following, let us consider the two classes of solutions given,
respectively, by Eqs.(\ref{class1}) and (\ref{class2}) separately.

\subsection{$y  \not= -1$}

In this case the solutions are given by Eq.(\ref{class1}). It can
be shown that the regularity conditions (\ref{cd1})-(\ref{cd3})
and the gauge one (\ref{cd4}) require
  \bq
  \lb{3.4}
  \alpha < 2, \;\;\; \Phi_{0} = 0, \;\;\;
  S_{0} = \frac{2}{2-\alpha}e^{\Psi_{0}}.
  \eq
On the other hand, using the transformations (\ref{rescaling}), we
can further set $\Psi_{0} = 0$. Then, we can see that in this case
there is only one free parameter,  $\alpha$, which characterizes
the kinds of self-similarity. Corresponding to Eq.(\ref{3.4}) and
the gauge choice $\Psi_{0} = 0$ we find that
  \bqn
  \lb{3.5}
   R &=& g^{\alpha\beta}R_{\alpha\beta} =
   g^{\alpha\beta}\phi_{,\alpha}\phi_{,\beta} =
   \frac{1}{4l^{2}(-t)^{2}},\nb\\
  \theta_{l} &=& \frac{f e^{\alpha x/2}}{2l^{2}r(-t)^{1/2}}
  \left[(2- \alpha)(-t)^{1/2} - r^{(2-\alpha)/2}\right],\nb\\
  \theta_{n} &=& - \frac{g e^{\alpha x/2}}{2l^{2}r(-t)^{1/2}}
  \left[(2- \alpha)(-t)^{1/2} + r^{(2-\alpha)/2}\right].
  \eqn
 From the above expressions we can see that the spacetime is
singular on the hypersurface $ t = 0$. One the other hand, we also
have $\theta_{n} < 0$ for any given $t \le 0$ and $r \ge 0$, while
$\theta_{l}$ has the properties
  \bq
  \lb{3.6}
  \theta_{l} = \cases{> 0, & $ r < r_{AH}(t)$,\cr
      = 0, & $ r = r_{AH}(t)$,\cr
      < 0, & $ r > r_{AH}(t)$,\cr }
  \eq
where
  \bq
  \lb{3.7}
  r_{AH}(t) \equiv \left[\left(2-\alpha\right)
  (-t)^{1/2}\right]^{2/(2-\alpha)}.
  \eq
 From the above expression we can see that the quantity
$\theta_{l}\theta_{n}$  is negative in the region where $t \le 0$
and $r < r_{AH}(t)$, which will be referred to as Region $II$,
while in the region where $t \le 0$ and $r > r_{AH}(t)$, which
will be referred to as Region $I$,  it  changes its sign and
becomes positive, $\theta_{l}\theta_{n} > 0$ [cf. Fig. 1]. Thus,
all the rings of constant $t$ and $r$ are trapped in Region $I$
but not in Region $II$. Note that the spacetime is singular on the
hypersurface $t = 0$ in Region $I$, which forms the up boundary of
the spacetime. From Eq.(\ref{3.5}) we can also see that the scalar
field is timelike in both of the two regions. Then, we can
consider Region $I$ as the interior of a black hole, which is
formed from the gravitational collapse of the scalar field in
Region $II$. The hypersurface
  \bq
  \lb{3.8}
  r = r_{AH}(t),
  \eq
represents an apparent horizon \cite{Hay94}.

  \begin{figure}[htbp]
  \begin{center}
  \label{fig1}
  \leavevmode
   \epsfig{file=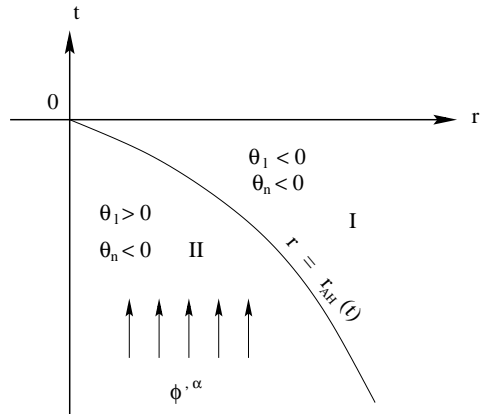,width=0.35\textwidth,angle=0}
  \caption{The spacetime in the ($t, r$)-plane for the solutions
  given by Eq.(\ref{class1}) with $\alpha < 2$. It
  is regular at the axis $r = 0$, but singular on the spacelike
  hypersurface $t = 0$.
  In Region $I$  all the rings of
  constant $t$ and $r$ are trapped, but not in Region $II$.
  The hypersurface $r = r_{AH}(t)$
  represents apparent horizon. }
  \end{center}
  \end{figure}

To study the above solutions further, let us introduce two new
coordinates $\bar{t}$ and $\bar{r}$ via the relations,
  \bq
  \lb{3.9}
  t = - \frac{1}{4}\left(-\bar{t}\right)^{2},\;\;\;
  r = \left(\frac{2-\alpha}{2}\bar{r}\right)^{2/(2-\alpha)},
  \eq
then we find that in terms of $\bar{t}$ and $\bar{r}$ the metric
takes the form,
  \bq
  \lb{3.10}
  ds^{2} = \frac{l^{2}}{4}\left(- \bar{t}\right)^{2}
           \left(d\bar{t}^{2} - d\bar{r}^{2} -
           \bar{r}^{2}d\theta^{2}\right),
  \eq
which shows that the solutions are actually conformally flat. On
the other hand, from Eqs.(\ref{3.8}) and (\ref{3.9}) we can see
that the apparent horizon located on the hypersurface $r =
r_{AH}(t)$ in the $(t,r)$-plane is mapped to the one $\bar{t} = -
\bar{r}$ in the $(\bar{t}, \bar{r})$-plane, from which it is
clearly that this surface is null and the corresponding Penrose
diagram is given by Fig. 2. 

  \begin{figure}[htbp]
  \begin{center}
  \label{fig2}
  \leavevmode
   \epsfig{file=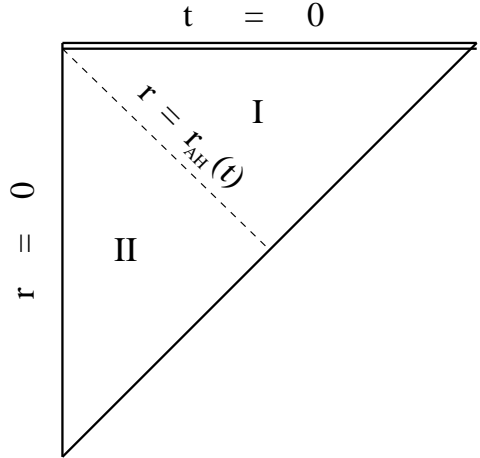,width=0.35\textwidth,angle=0}
  \caption{The Penrose diagram for the solutions given  by
  Eq.(\ref{class1}) with $\alpha < 2$.  The  spacetime is singular
  on the double line $t = 0$. All the rings of constant $t$ and $r$ are
  trapped in Region $I$ but not in Region $II$.
  The hypersurface $r = r_{AH}(t)$ is a null surface and represents
  the apparent horizon.}
  \end{center}
  \end{figure}

It should be noted that the above analysis is valid only for
$\alpha < 2$, as Eq.(\ref{3.4}) tells. When $\alpha = 2$, we have
  \bq
   \lb{3.11}
   {\cal{R}} =  S_{0}(-t)^{1/2},\;\; (\alpha = 2),
  \eq
from which we can see that the conditions (\ref{cd1})-(\ref{cd2})
cannot be satisfied. As a matter of fact, we always have
${\cal{R}}_{,\alpha}{\cal{R}}^{,\alpha} > 0$, and the
corresponding solution is Kantowski-Sachs like \cite{Kramer80},
and may be considered as representing cosmological models.

When $\alpha > 2$, we find that
  \bq
  \lb{3.12}
  {\cal{R}} =  \frac{S_{0}(-t)^{1/2}}{r^{(\alpha - 2)/2}}
  = \cases{0, & $r \rightarrow \infty$,\cr
       \infty, & $r = 0$.\cr}
  \eq
Thus, now the axis is located at $r = \infty$. Introducing the new
coordinate $\bar{r} = 1/r$, we find that the corresponding metric
in terms of $\bar{r}$ takes the form
   \bq
   \lb{metric2}
   ds^{2} = l^{2}\left\{e^{2\Phi(\bar{x})}d\bar{t}^{2}
   - e^{2\Psi(\bar{x})}d\bar{r}^{2} -
   \bar{r}^{2}S^{2}(\bar{x})d\theta^{2}\right\},
   \eq
where $\bar{t} = t$, and
  \bqn
  \lb{class1b}
  \Phi(\bar{x}) &=& \Phi_{0},\;\;\;
  S(\bar{x}) = S_{0}e^{-\bar{\alpha} \bar{x}/2},\;\;\;
  \Psi(\bar{x}) = -\frac{1}{2}\bar{\alpha} \bar{x} + \Psi_{0},
  \nb\\
  \bar{x} &\equiv& \ln\left(\frac{\bar{r}}
  {(-\bar{t})^{1/\bar{\alpha}}}\right),
  \;\;\;
  \bar{\alpha}  < 2,
  \eqn
with $\bar{\alpha} \equiv 4 - \alpha$. Dropping the bars in the
above equations, we can see that they are exactly the solutions
given by Eq.(\ref{class1}) with $\alpha < 2$. Therefore, {\em
solutions (\ref{class1b}) with $\bar{\alpha} < 2$ or ($\alpha
 > 2$) describe the same spacetimes as the ones given by
(\ref{class1}) with $\alpha < 2$}.

\subsection{$y = -1$}

In this case the solutions are given by Eq.(\ref{class2}). The
transformations (\ref{rescaling}) and the gauge condition
(\ref{cd4}) enable us to set $\Phi_{0} = \Psi_{0} = 0,\; S_{0} =
1$. Then, the metric  takes the form
  \bq
  \lb{3.13}
  ds^{2} = l^{2}\left\{dt^{2} - e^{2(1-\alpha)x}dr^{2} -
   (-t)^{2/\alpha}d\theta^{2}\right\},\;\; (\alpha > 1),
  \eq
from which we can see that the solutions, similar to the case
$\alpha = 2$ given in the last subsection,  are also
Kantowski-Sachs like and do not satisfy the conditions (\ref{cd1})
and (\ref{cd2}). The spacetime is singular at $ t = 0$, which can
be seen, for example, from the expression
  \bq
  \lb{eq20}
  R = \phi_{,\alpha}\phi^{,\alpha} =
  \frac{\alpha -1}{ l^{2}(\alpha t)^{2}}.
  \eq

\section{Conclusion}

In this paper, we found all the solutions of the
Einstein-massless-scalar field equations with kinematic
self-similarity of the second kind in the $(2+1)$-dimensional
spacetimes with circular symmetry, which consist of two classes
given, respectively, by Eqs.(\ref{class1}) and (\ref{class2}).

It was shown that the solutions given by (\ref{class1}) with
$\alpha < 2$ represent gravitational collapse of the scalar field,
and the collapse always forms black holes. The solutions
Eqs.(\ref{class1}) with $\alpha > 2$ describe the same spacetimes
as those with $\alpha < 2$, after replacing $\alpha, \; r$ by
$\bar{\alpha}, \; \bar{r}$, respectively, where $(\bar{\alpha},
\bar{r})=(4 - \alpha, 1/r)$. It was also shown that the solution
Eq.(\ref{class1}) with $\alpha = 2$ and the ones given by
Eq.(\ref{class2}) are all Kantowski-Sachs solutions but in
$(2+1)$-dimensional spacetimes. These solutions may represent
cosmological models with a spacetime singularity located on a
spacelike hypersurface.

It is somehow surpprise that the solution with second
kind self-similarity given by Eqs.(\ref{class1}) and (\ref{3.4})  is identical 
to the one found by Cl\'ement and Fabbri \cite{CF01} for the same type
of fluid but with the self-similarity of the first kind. 
This shows that a spacetime can have two different kinds of
self-similarities, that is, there exist two vector fields, say,
${\xi^\mu}_{(1)}$ and ${\xi^\mu}_{(2)}$, where ${\xi^\mu}_{(1)}$ describes the
self-similarity of the first kind, while ${\xi^\mu}_{(2)}$ the second kind.
This happens when the spacetime has high symmetry.

We also found, without demonstrations, that solutions with
self-similarity of the zeroth kind to the Einstein-massless-scalar
field equations in the spacetime considered here do not exist.

Finally, we would like to note that Ida recently showed that a
(2+1)-dimensional gravity theory which satisfies the dominant
energy condition forbids the existence of black holes
\cite{Ida00}. This result does not contradict with ours obtained
here, as in the present case the surfaces of apparent horizons of
the black holes are degenerate \cite{Hay94}.

\section*{Acknowledgments}

The financial assistance from UERJ (JFVdaR) and FAPERJ/UERJ (MFAdaS) is
gratefully acknowledged.


\end{document}